# Stablecoins and the Emerging Hybrid Monetary Ecosystems

Hongzhe Wen, Washington University in St. Louis (hongzhe.w@wustl.edu)

Songbai Li, University of California San Diego (sol026@ucsd.edu)

R.S.M. Lau, Hong Kong University of Science and Technology (rlau@ust.hk)

Jamie Zhang, Drexel University (jz837@drexel.edu)

**ABSTRACT**

With market capitalization exceeding USD250 billion by mid-2025, stablecoins have evolved from a crypto-focused innovation into a vital component of the global monetary structure. This paper identifies the characteristics of stablecoins from an analytical perspective and investigates the role of stablecoins in forming a hybrid monetary ecosystem where public (fiat, CBDC) and private (USDC, USDT, DAI) monies coexist. Through a number of econometric analysis models, we find that stablecoins maintain strong peg stability, while each type exhibiting distinctive responses to market variables such as trading volume and capitalization depending on the mechanisms behind. We also introduce a hybrid system design that proposes a two-layer structure, which private stablecoin issuers are backed by central bank reserves, ensuring uniformity, security, and programmability. This model takes advantages of both decentralized finance and payment innovation, while utilizing the Federal Reserve's institutional trust. A case study on the SVB-USDC de-peg event in 2023 illustrates how such a hybrid system could have prevented panic-induced instability through transparent reserves, secured liquidity, and interoperable assets. Through examination of the Dybvig model and simulation, we conclude that a hybrid monetary model not only enhances financial inclusivity, scalability, and dollar utility in digital ecosystems, but it also strengthens systemic resilience, offering a credible blueprint for future digital dollar architectures.

## 1. Introduction

Stablecoins are digital currencies designed to maintain a stable value relative to certain assets such as fiat money. The vast majority of stablecoins are now backed by U.S. dollar. Their ability to facilitate peer-to-peer (P2P) or cross-border transactions instantly with little costs, among other functions, helps position stablecoins as a disruptive technology of the existing payments system. Stablecoins have rapidly evolved into a "killer app" bridging the crypto ecosystem and blockchain technology with traditional finance market and real-world applications. In just a few short years, the total market capitalization of stablecoins has grown to USD250 billion by mid-2025. Trading volume of stablecoins now exceeds those of Bitcoin and Ethereum, the most well-known applications of cryptocurrency.

Yet this dramatic rise of stablecoins also raises questions for the future of the existing monetary systems: If billions of people can switch to privately issued digital dollars without any limitation, how is it going to affect the traditional fiat currencies, central banks' control over money, and financial stability? As a result, many see the disruptive nature of stablecoins as they may erode the common monetary policy tools that central banks around the world have been using to control the money supply due to decentralized issuance and easy cross-border money flow of stablecoins.

These questions place stablecoins at the center of controversies and therefore lead to the need of introducing a new hybrid monetary ecosystem, where public and private money interact in unprecedented ways. There are many discussions on central bank digital currency (CBDC) in recent years, but with the passing of the GENIUS Act in the United States in July 2025, it becomes obvious that any CBDC will have a very limited application going forward. The focus should be on how to develop a hybrid monetary architecture to integrate stablecoins into the existing monetary base while addressing the volatile nature of cryptocurrencies.

This paper examines the stabilities of stablecoins and then introduces a hybrid monetary ecosystem in which stablecoins and fiat currencies coexist and integrate. Our proposed hybrid monetary ecosystem design, which private stablecoin issuers are backed by central bank reserves, introduces new operational and architectural elements to improve scalability, cross-border interoperability, and systemic resilience.

We will consider quantitative analysis results (summary statistics, volatility and correlation metrics and time-series test) to identify the characteristics of stablecoins to support our proposed monetary model. A simulation model with reference to the SVB-USDC de-peg event in 2023 will help illustrate that a hybrid monetary model not only enhances financial inclusivity, scalability, and dollar utility in digital ecosystems, but it also strengthens systemic resilience, offering a credible blueprint for future digital dollar architectures.

## 2. Background

Efficient and reliable monetary ecosystems are essential to facilitate all kinds of trade and everyday financial transactions. Nowadays, these ecosystems consist of a complex network of interactions among traditional financial institutions, central banks, digital currencies, and emerging technologies like blockchain and open finance platforms.

### *2.1. Stablecoins and monetary infrastructure*

Stablecoins are tokenized cash issued by private companies on blockchain (e.g., Ethereum) that are primarily used for trading crypto assets, transacting in goods and services, insulating against local currency instability, and sending payments across borders [1]. These stablecoins generally fall into three design categories: fiat-collateralized, crypto-collateralized, and algorithmic [2]. The most prevalent ones are fiat-backed stablecoins such as USDT and USDC, which aim to maintain 1:1 parity with the U.S. dollar by holding reserves in cash, bank deposits, or short-term treasuries. Their stability relies on both market arbitrage and redemption mechanisms: when prices deviate from $1, arbitrageurs buy or redeem the token to restore the peg. In practice, these mechanisms function smoothly when reserves are credible, liquid, and transparently disclosed [3].

In contrast, crypto-collateralized stablecoins like DAI maintain price stability using overcollateralized smart contracts. Users lock up (as collateral) volatile assets (e.g., ETH or WBTC) in “vaults” to mint DAI, and the protocol includes automated liquidation mechanisms and stability fees to manage supply and demand [4]. These systems are more decentralized but rely on strong collateralization ratios to defend the peg. Algorithmic stablecoins, such as the now-defunct TerraUSD, attempt to maintain a peg through endogenous monetary policy, minting and burning supply based on price, but have proven prone to collapse under stress [5].

### *2.2. Central bank digital currencies (CBDCs)*

As private stablecoins have grown, some central banks worldwide have accelerated efforts to develop their own digital currencies. CBDC is a digital form of fiat money, issued and backed by a central bank, intended to serve as legal tender [6]. In essence, a retail CBDC would be a digital cash analog, which is a liability of the central bank that the public could hold directly, using wallets or accounts managed by the central bank or intermediaries [7].

Since 2020, People’s Bank of China, the central bank in China, has been piloting its retail CBDC, commonly called the e-CNY, in 23 provinces and major cities using a two-layer distribution network of commercial banks and payment platforms. By mid-2024, transaction volumes of e-CNY exceeded 7 trillion CNY (about USD1 trillion) across over 260 million wallet users, making it the largest CBDC in the world. Key features of e-CNY include offline peer-to-peer transfers, bank-backed wallet interchangeability, and programmable conditional payments, offering a real-world testbed for privacy, scalability, and cross-border use cases [8].

### *2.3. Global regulatory landscape*

By July 2025, most major financial centers, including the United States, European Union, Singapore, Switzerland, Hong Kong, and the United Arab Emirates, have adopted formal legal frameworks for fiat-backed stablecoins. These regulatory regimes have legitimized stablecoins as financial instruments and placed them under prudential supervision, typically requiring full reserve backing, redemption guarantees, and institutional licensing. This shift has helped reframe stablecoins as programmable financial infrastructure rather than speculative crypto assets.

However, regulatory inconsistencies remain across different jurisdictions. These inconsistencies are caused by various redemption timelines (e.g., 90 days in Hong Kong [9] vs. 1 day in the United States [10]), transaction caps (e.g., the €200 million daily limit under MiCA in Europe [11]), and differences in supervisory authorities (banking vs. securities regulators). Such divergences introduce operational frictions for cross-border usage, custody localization, and compliance scalability, particularly for stablecoin issuers seeking to operate across multiple legal domains.

These frictions are not just transitional, they are structural. As such, they directly shape the system architecture proposed in this paper. Rather than assuming regulatory harmonization, our hybrid model is explicitly designed to accommodate this heterogeneity through interoperable token standards, verifiable reserve structures, and programmable redemption protocols. These features enable stablecoin parity, liquidity, and compliance consistency across fragmented jurisdictions.

These regulatory divergences are not negligible, rather they define the operational constraints our proposed architecture must address. In section 5, we introduce the hybrid monetary model designed to accommodate these frictions and ensure systemic interoperability.

## 3. Data analysis results and insights

Before we introduce our proposed hybrid monetary ecosystem, we would like to present some data analysis results and insights we have discovered for the adoption and stability of stablecoins. Our analysis includes USDC, USDT, and DAI (representing more than 90% of the total market capitalization). For detailed descriptions of the analysis methodologies and the simulation model, please refer to the Appendix.

### *3.1. Stablecoin adoption trends*

Empirical data illustrate both the scale and the shifting patterns of stablecoin use. According to the data in CoinGecko, the market capitalization of stablecoins climbed to around USD240 billion as of May 2025. This expansion has been driven by user demand for stable value in crypto trading, decentralized finance (DeFi) applications, and as a dollar substitute in various economies.

The usage trends are verified by volume data: Stablecoins now routinely handle daily transaction volumes in the tens of billions of dollars across exchanges and on-chain transfers. Our analysis in Table 1 shows that on average, USDT alone has USD30.7 billion in average daily trading volume, far exceeding that of USDC and DAI. This reflects USDT's growing role as the base trading pair on many cryptocurrency exchanges and a popular tool for moving funds between exchanges or into fiat money.

**Table 1**
Descriptive statistics of DAI, USDC, and USDT (in mid-2025).

| Coin | Price Mean | Price SD | Market Cap Mean | Market Cap SD | Volume Mean |
|---|---|---|---|---|---|
| DAI | 1.001513 | 0.00678 | 4,277,854,905 | 2,568,008,276 | 283,750,214 |
| USDC | 1.000896 | 0.00397 | 23,512,851,601 | 19,538,081,668 | 3,781,436,313 |
| USDT | 1.000650 | 0.01418 | 37,637,875,312 | 44,352,804,356 | 30,717,253,060 |

In terms of user adoption, blockchain data (per industry reports) show that the number of addresses holding stablecoins and the transaction counts have continued to grow when speculative crypto trading activity weakened, indicating stablecoins have uses beyond just trading. During the crypto bear market in 2022, exchange trading volumes fell tremendously, but on-chain stablecoin transaction counts remained strong or even grew [12], suggesting people continued using stablecoins for payments, on chain financial savings, or as a haven, during that period. Another trend is the growing acceptance of stablecoins in mainstream fintech and commerce. For example, some payment processors now support stablecoin settlement for merchants to increase the efficiency of cash circulation within the supply chain. In 2023, PayPal launched PYUSD, allowing merchants to receive cash faster [13].

### *3.2. Peg stability and volatility analysis*

In Fig. 1, daily price data (sample through March 2025) show that all three stablecoins maintained average prices extremely close to $1.00 except during the first few days after launch. Among the selected stablecoins, USDC had the highest fidelity, with an average absolute peg deviation of only 0.19% and a standard deviation of 0.40%. DAI's price deviated by 0.32% on average with a standard deviation of 0.68%, while USDT exhibited 0.28% average deviation and the largest dispersion (s.d. 1.42%) due to a few outlier events. USDT also achieved the peg on the greatest proportion of days after launched (62.6% of days at exactly $1, vs. 57.8% for USDC and 51.6% for DAI).

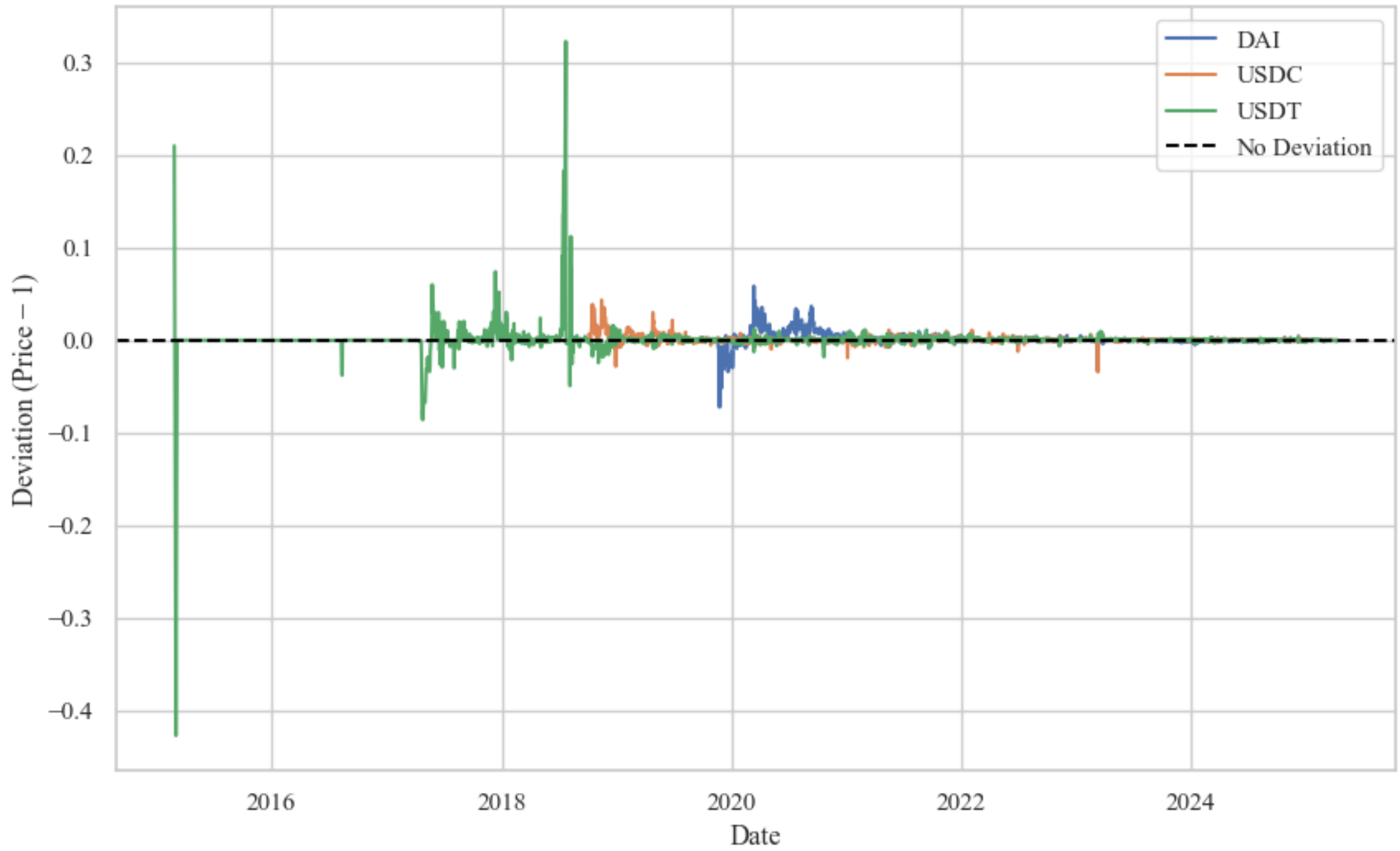

**Fig. 1.** Stablecoin peg deviation over time.

However, USDT experienced the single largest deviation in the sample. As shown in Table 2, at one point, USDT was trading 42.75% away from $1 during a severe early incident, whereas DAI's worst case was a 7.25% deviation and USDC's 4.35%. These extremes were short-lived, and typically the coins stayed within a much tighter range around $1.

**Table 2**
Result of deviation and peg calculations.

| Coin | Avg. Abs. Deviation | Std. Dev. (Price) | Max Deviation | Off-Peg Days (%) | Longest Peg Duration |
|---|---|---|---|---|---|
| USDC | 0.001901 (~0.19%) | 0.003971 (~0.40%) | 0.043465 (4.35%) | 42.23% | 38 |
| USDT | 0.002757 (~0.28%) | 0.014176 (~1.42%) | 0.427479 (42.75%) | 37.37% | 70 |
| DAI | 0.003216 (~0.32%) | 0.006780 (~0.68%) | 0.072464 (7.25%) | 48.38% | 162 |

Notably, DAI once remained at the peg for 162 consecutive days (its longest run without significant deviation), compared to 70 days for USDC and 38 days for USDT, indicating that DAI can sustain long periods of stability under past market conditions.

During tranquil periods, deviations have been minor. All three stablecoins traded extremely close to $1 on a day-to-day basis, with small fluctuations on the order of a few tenths of a percent. To quantify the short-term volatility, we examine rolling standard deviations of price. Table 3 reports the average weekly and monthly volatility for each coin. USDC again displays the least volatility, with a 7-day rolling standard deviation averaging only 0.16% and a 30-day (monthly) volatility of 0.20%. DAI's volatility is slightly higher (0.20% weekly and 0.25% monthly), while USDT shows the highest volatility (0.23% weekly and 0.34% monthly on average).

**Table 3**
Rolling volatility of stablecoin prices.

| Coin | Avg. Weekly Volatility ($\sigma_7$) | Avg. Monthly Volatility ($\sigma_{30}$) |
|---|---|---|
| USDC | 0.16% | 0.20% |
| USDT | 0.23% | 0.34% |
| DAI | 0.20% | 0.25% |

*Average 7-day and 30-day rolling standard deviations (σ) of daily price illustrate the low inherent volatility of stablecoin exchange rates. All values are in percentage terms.*

This ranking aligns with the intuitive stability order: USDC < DAI < USDT in terms of typical short-run price fluctuation magnitude. Notably, all these volatility figures are very low in absolute terms, well under 0.5% in normal conditions, highlighting that these stablecoins generally trade in an extremely tight range around their $1 target.

## *3.3. Time series and econometric analysis*

To assess the stationarity properties of these price series and their interactions, we conduct unit root tests and time-series modeling. Augmented Dickey-Fuller (ADF) tests firmly reject the null hypothesis of a unit root for each stablecoin's price series, confirming that these series are stationary. The ADF test statistic for USDC is –8.94 ($p < 0.001$), far below the 1% critical value (–3.43), indicating that USDC's price does not follow a random walk but instead mean-reverts around the peg.

Similarly, USDT and DAI show ADF t-statistics of –6.07 and –3.52, respectively (both significant at the 5% level). This means any deviation of price is temporary, and as designed, prices tend to revert toward $1 rather than wandering arbitrarily. Intuitively, the presence of arbitrage and redemption mechanisms (for USDC/USDT) or collateral incentives (for DAI) creates a restoring force that pulls the price back to parity, yielding statistically mean-reverting behavior.

## *3.4. Regression analysis of peg deviations*

Finally, we quantitatively link stablecoin peg deviations to market drivers like trading volume, market capitalization changes, and broader macroeconomic conditions. We estimate separate regression models for each coin and macroeconomic indices, using daily data. The dependent variable in each model is the peg deviation using the daily percentage deviation from $1. Key independent variables include trading volume and market cap, as well as the previous day's peg deviation (to capture persistence) and lags of volume and market cap. Table 4 summarizes the estimated coefficients.

**Table 4**

Regression outcomes for drivers of daily peg deviation.

| Variable | DAI (coefficient) | USDC (coefficient) | USDT (coefficient) |
|---|---|---|---|
| Constant | $8.25 \times 10^{-4}$ (***) | $4.88 \times 10^{-4}$ (***) | $4.01 \times 10^{-4}$ (n.s.) |
| Total Volume | $-8.91 \times 10^{-13}$ (*) | $-1.33 \times 10^{-14}$ (n.s.) | $-4.77 \times 10^{-16}$ (n.s.) |
| Market Cap | $9.61 \times 10^{-13}$ (n.s.) | $9.71 \times 10^{-13}$ (***) | $4.21 \times 10^{-12}$ (***) |
| PegDev (lag1) | 0.783 (***) | 0.709 (***) | 0.382 (***) |
| Volume (lag1) | $1.39 \times 10^{-12}$ (**) | $1.79 \times 10^{-14}$ (n.s.) | $3.52 \times 10^{-15}$ (n.s.) |
| M.Cap (lag1) | $1.11 \times 10^{-12}$ (n.s.) | $-9.77 \times 10^{-13}$ (***) | $-4.21 \times 10^{-12}$ (***) |

*Coefficient estimates (with significance in parentheses) from OLS models of each coin's deviation from $1.00. The independent variables are contemporaneous daily trading volume and market cap, plus one-day lag deviation, volume, and market cap. (Significance codes: *** $p < 0.001$, ** $p < 0.01$, * $p < 0.05$; n.s. = not significant.)*

Several clear patterns emerge from these regression analyses. Peg deviations are highly persistent from one day to the next, especially for DAI and USDC. The coefficient on the lagged deviation is 0.783 for DAI and 0.709 for USDC (both $p < 0.001$), meaning a deviation does not fully correct within a day, roughly 70% to 78% of it carries over to the next day.

This aligns with our earlier finding that DAI can take a few days to mean-revert. USDT's lagged deviation coefficient is lower (0.382), indicating that USDT's peg deviations tend to correct more rapidly (less than 40% of a shock remains after one day), consistent with the very quick re-pegging that we observe empirically for USDT in most cases.

Trading volume has a significant impact on DAI's peg, but not on USDC or USDT. For DAI, the coefficient on contemporaneous volume is $-8.91 \times 10^{-13}$ ($p \approx 0.046$), which is statistically significant. This negative sign implies that on days when DAI's trading volume is high, DAI's price deviation is slightly smaller (i.e., closer to $1).

In practical terms, if DAI's volume were to increase by, say, USD1 billion in a day (a considerably large jump, given DAI's mean daily volume of merely USD280 million), the model predicts the DAI price would be about 0.09% closer to the peg (i.e., $(-8.91 \times 10^{-13}) \times (1 \times 10^{9}) \approx -8.91 \times 10^{-4}$). While small, this effect supports the intuition that higher trading activity helps enforce DAI's peg via arbitrage.

The lag volume for DAI has a positive coefficient ( $1.39 \times 10^{-12}$, $p < 0.01$), suggesting that an unusually high-volume day might be followed by a widening of DAI's deviation the next day, which possibly because the initial high-volume event was a response to a shock that still has some residual effect.

In contrast, USDC and USDT show no significant direct relationship between daily volume and peg deviation. Their volumes (which are an order of magnitude larger than DAI's) do not exhibit a measurable impact on price stability in the regression – likely because both coins have well-designed institutional mechanisms (and ample liquidity) that keep the price at $1 regardless of trading frenzy.

Market capitalization changes are significant for the fiat-backed coins. For USDC and USDT, the current market cap term is positive and highly significant ($p < 0.001$), while the one-day lag market cap term is negative and of similar magnitude (also significant). This indicates a short-lived effect of supply changes on the peg. Specifically, when a large amount of USDC is issued (increasing its market cap), USDC tends to trade at a slight premium on that same day. From the result, the USDC coefficient of $9.71 \times 10^{-13}$ implies that if USDC's market cap were to rise by USD1 billion, the immediate deviation might increase by ~0.10%. This could reflect that new issuance often responds to heightened demand (USDC trading above $1 triggers the creation of new coins, which initially still leaves a small premium until enough supply comes in).

However, the negative lag effect ($-9.77 \times 10^{-13}$) suggests that by the next day, the deviation is pushed back down by a similar amount, as the increased supply fully meets demand and the price returns to parity. USDT shows an analogous pattern: A positive same-day effect of market cap ($4.21 \times 10^{-12}$, $p < 0.001$) and a nearly equal negative next-day effect ($-4.21 \times 10^{-12}$), again point to very short-term impacts of supply shocks that reverse within a day.

In simpler terms, minting or redeeming a large volume of USDC/USDT can momentarily knock the price off $1, but the deviation corrects almost immediately. DAI's market cap variables, on the other hand, were not significant, implying that changes in DAI supply (which are usually slower and governance-mediated) did not have a direct day-to-day effect on its price in the sample.

It is also worth noting that broader macroeconomic variables (interest rates, money supply, inflation) included in the preliminary models do not show any statistically significant influence on these peg deviations at a daily frequency. The stability of these coins seems to be primarily a function of crypto-market dynamics and internal mechanisms rather than macroeconomic conditions.

## 4. System Design: Architecture of a Hybrid Monetary Ecosystem

In this section, we will first discuss the difference between a hybrid CBDC model and our proposed hybrid monetary ecosystem facilitated by stablecoins. We will then highlight the features that our proposed model can effectively address those limitations of the hybrid CBDC model.

### *4.1. Comparison to a hybrid CBDC model*

There are many discussions on a hybrid CBDC model. Our proposed hybrid monetary architecture diverges from several influential public and academic models of digital currency design. In recent years, central banks and scholars have outlined various frameworks for integrating digital assets into the monetary base. Our approach, while aligning with the stability and transparency goals of those already proposed by others, introduces new operational and architectural elements to improve scalability, cross-border interoperability, and systemic resilience.

One prominent policy framework is the Federal Reserve's intermediated CBDC model [14], which envisions a central bank–issued retail CBDC distributed through regulated intermediaries. While this design ensures legal tender status and systemic oversight, it imposes substantial technical and institutional overhead on central banks, and does not directly leverage the existing adoption of private stablecoins. In contrast, our model treats stablecoins as composable building blocks, embedding them into a broader programmable ecosystem while preserving the Federal Reserve's control over monetary settlement and liquidity.

The IMF also proposed a synthetic CBDC [15] with the idea of private entities issuing coins fully backed by central bank reserves. Our hybrid model draws from this logic but extends it in three key ways: (i) we introduce programmable interoperability across issuers; (ii) we simulate system-wide peg stability under real-world stress; and (iii) we propose an architectural role for smart contracts and decentralized finance (DeFi) platforms, which are features absent from the IMF's original proposal.

From the academic side, Gorton and Zhang [16] argue that unregulated stablecoins resemble 19th-century wildcat banks and require either conversion into public money or strict collateral constraints. Our hybrid architecture responds directly by enforcing 100% reserve backing in central bank assets and by integrating a compliance layer that satisfies both AML/KYC and consumer protection expectations.

Furthermore, our proposed model incorporates insights from the narrow banking literature [17,18], in which banks issue only fully backed liabilities, eliminating maturity transformation risk. We embed this principle within a tokenized infrastructure, transforming narrow banking into a programmable digital currency layer without forgoing the benefits of composability, smart contract utility, or interoperability across jurisdictions.

Finally, Auer and Böhme [19] and the Bank for International Settlements (BIS) have proposed modular CBDC frameworks, emphasizing technical architectures like "retail/wholesale layering" and "intermediated identity channels." Our proposed model operationalizes these concepts in a live crypto-financial context by showing how existing stablecoin rails can be federated across jurisdictions while preserving privacy, enabling compliance enclaves, and integrating with on-chain governance tools.

In summary, our model design contributes to this growing literature by offering a simulation-evaluated, compliance-conscious, and digitally native architecture that preserves the creditworthiness of public money while extending its reach and programmability through private innovation.

### *4.2. The proposed hybrid monetary ecosystem design*

To further enhance the stability of stablecoins, the proposed hybrid monetary system is a two-layer architecture combining private-sector innovation at the user interface and central bank support at the core:

- Layer 1 – Core CBDC infrastructure (public): The Federal Reserve operates the wholesale/retail CBDC ledger and grants interfaces to regulated intermediaries. It supplies the default-free settlement asset while leaving retail customer relationships to the private sector.
- Layer 2 – Private tokens and deposits (private): Banks or fintech issuers

tokenize deposits and circulate stablecoins that are redeemable 1:1 for CBDC. Reserve assets are held as CBDC balances or Fed-custodied Treasuries, making issuers functionally narrow banks. Interoperability is achieved through shared messaging and redemption protocols.

At a high level, regulated private entities (stablecoin issuers and banks) would issue digital USD tokens and provide payment services, while the Federal Reserve and government establish or adopt the common foundation (rules, settlement infrastructure, and backstop).

As show in Fig. 2, this structure mirrors an "intermediated" model of CBDC but extends it by treating existing fiat-backed stablecoins as a feature rather than a competitor. The design rests on several pillars as explained later.

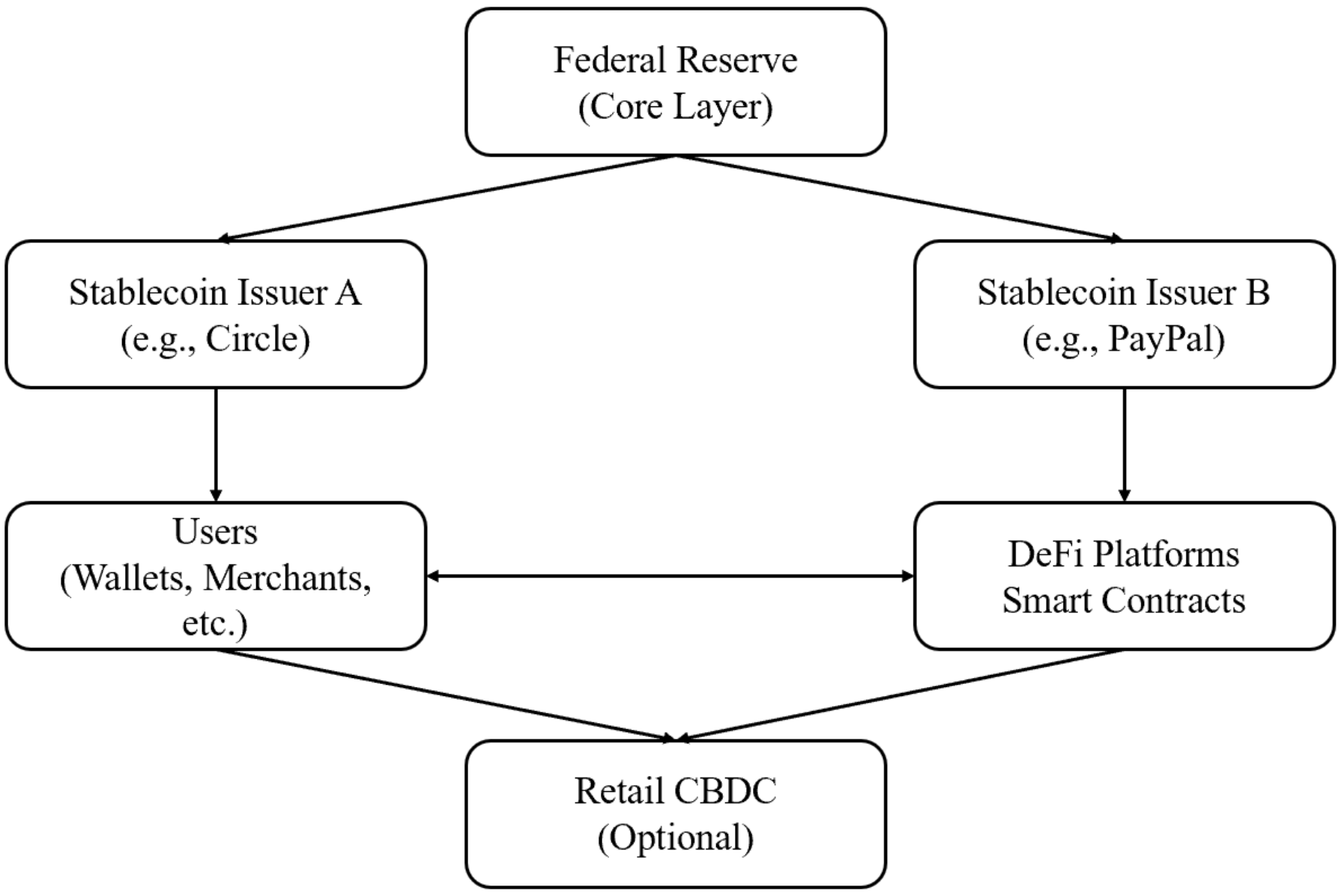

**Fig. 2.** Hybrid monetary system design.

## *4.3. Private digital dollar issuers*

In our proposed model, fintech companies or banks will issue tokenized USD liabilities (stablecoins) that are fully backed by high-quality, liquid assets or actual fiats. Each stablecoin token represents a claim to $1 in reserves. Users can hold and transfer these tokens on various blockchain networks, or called public chains (Ethereum, Solana, etc.), enabling instant P2P payments, smart contract integration in DeFi, and more. These issuers will operate under charters/licenses that meet strict compliance requirements.

In practice, this means firms like Circle (issuer of USDC), Tether (issuer of USDT), or other non-bank payment firms (e.g., PayPal with PYUSD) could become "Permitted Payment Stablecoin Issuers" as defined in new legislation. Traditional banks could also issue their own tokenized deposits, but the model focuses on interoperable stablecoins that any user can accept, rather than each bank creating a silo stablecoin and compete.

To ensure zero-delay USD conversion, issuers will connect directly into FedNow rails: any on-chain redemption request triggers a same-second FedNow credit to the user's bank account. This binding commitment eliminates common "cash-out" issues in private stablecoins today.

### *4.4. 100% reserve backing (synthetic CBDC model)*

To ensure uniform value and eliminate liquidity (similar to a bank run) risk, each stablecoin issuer in the system need to hold reserves equal to 100% of their outstanding tokens in safe assets (USD cash or equivalents). The optimal arrangement, often called a "synthetic" CBDC, is for issuers to deposit these reserves directly in the Fed account.

The stablecoin effectively becomes a narrow-bank liability fully backed by central bank money. This gives the tokens or stablecoins the same creditworthiness as CBDC (since the backing funds are a claim on the Federal Reserve) while still allowing private issuance. If direct Fed access for non-banks is initially unattainable, a similar effect can be achieved by holding reserves in short-term Treasury bills or insured bank deposits under trust. In the end, the goal is to minimize liquidity risk in reserves for stablecoins.

### *4.5. Federated networks and interoperability*

To prevent fragmentation and competition, the system requires interoperability among all official stablecoin tokens and with other forms of USD. A stablecoin issued by one licensed entity is fungible with another's and with traditional dollars. Achieving this requires common technical and legal standards; for example, standardized messaging formats and APIs to form bridges between different platforms, which enable transfer and swap tokens easily.

An interoperable design could allow a user with USDC tokens to easily send funds to someone who uses, for example, a PayPal PYUSD wallet, or pay directly into a merchant's bank account, which the underlying infrastructure would convert and route the payment, much like different banks clear checks at par nowadays.

The Federal Reserve could facilitate this by extending its payment systems (FedNow or FedWire) to support stablecoin settlement, e.g., allowing banks to settle stablecoin redemption flows in real time. In addition, industry consortia might develop common protocols or bridges to connect multiple blockchains, so stablecoin liquidity is not isolated on one chain.

### *4.4. Role of a retail CBDC*

The model does not necessarily require the Federal Reserve to issue a retail CBDC immediately but leaves space for one as a complementary option. The Federal Reserve could proceed with a limited retail CBDC aimed at specific use cases (for example, as a public option for those who prefer holding a direct claim on the central bank) or only available when off peg happens to stablecoins. This CBDC would function alongside stablecoins: Think of it as the digital equivalent of physical cash, whereas stablecoins are more like digital demand deposits issued by private institutions.

In our hybrid design, the CBDC and stablecoins coexist and are interchangeable. A user might hold some balance in CBDC for its sovereign guarantee, and some in a private stablecoin wallet for specialized services, such as DeFi lending/borrowing or international commerce purposes. These wholesale CBDC tokens would settle instantaneously in seconds and 24/7 with almost no marginal cost, mirroring the sub-cent fee structure observed in retail CBDC pilots such as the digital euro (€0 per user transaction) and Nigeria's eNaira (0 NGN per P2P transfer).

As shown in Table 5, this capability matches with, and could be layered atop, the FedNow real-time gross settlement rail (seconds; $0.045 per transfer), creating a unified, blockchain-compatible payment base. Comparing to other payment infrastructures, this hybrid model design is one of the most efficient means available.

**Table 5**
Transaction speed of different infrastructures in the system.

| **Payment Infrastructure** | **Settlement Time** | **Cost per Transaction** |
|---|---|---|
| ACH | Majority of payments settle next business day; standard ACH posts in 1–2 business days [20] | $0.26–$0.50 per transfer [21] |
| FedNow | Within seconds [22] | $0.045 per credit transfer |
| Visa | Authorization < 3s, funds to merchants 1-3 business days | 1.15%–3.15% of transaction value [23] |
| Ethereum | 5-20s to confirm (average 13 blocks) | $0.1839 average fee |
| Solana | ~400ms to finality | $0.0025 average fee |
| USD CDBC | Near instant [24] | $0 for individual |

**All crypto data from [25].*

While a retail CBDC offers consumers the same instant and near-zero-fee experience, it also raises privacy, disintermediation, and cybersecurity concerns that are less pronounced in the wholesale context. By confining CBDC tokens to the wholesale layer, banks and issuers can convert stablecoin liabilities into CBDC units on-chain without exposing end-users to those trade-offs: Preserving commercial bank funding while ensuring finality and minimal settlement risk. Ultimately, every tokenized dollar, whether retail stablecoin or wholesale CBDC, would need to clear back to a Fed liability, maintaining a single, centrally controlled monetary base that supports both innovation and financial stability.

### *4.5. Programmability and DeFi integration*

A key advantage of including private stablecoins is its rich ecosystem of decentralized finance (DeFi) and programmable money. The hybrid system would support smart contract functionality for the dollar. This means stablecoins can be used in automated agreements through smart contracts, e.g., release funds on delivery of goods, yield-generating decentralized lending, or micropayment streams by the second.

By design, the stablecoin tokens would be composable with existing DeFi protocols. In fact, USDC and DAI are already deeply embedded in lending platforms, decentralized exchanges, and so forth. The Federal Reserve or any regulators need not build this functionality themselves. Instead, they supervise the private service providers who innovate in it. The result is a programmable money layer on top of the dollar, integrating fintech innovation further into the system.

For businesses, this enables cost-effective payment logic and internal circulation (like supply chain payments that execute automatically upon IoT sensor triggers, etc.), and for consumers, it means access to a global financial marketplace using dollars in new ways (borrowing/lending, trading, automated investments) without currency exchange friction.

In the United States, this hybrid system helps with maintaining USD primacy to a further step by ensuring the dollar is the default currency for value in the future digital economies. That is, by providing official support for USD stablecoins, the United States can crowd out potential competitors, and such benefit maintained the same for all countries who adopt the system. Conversely, if no USD token were readily available, a crypto market might adopt a foreign stablecoin or even a volatile crypto as a unit of account, which ensures the stability of the hybrid system.

### *4.6. Monetary integrity and parity*

Crucially, this architecture preserves monetary integrity – all dollars, whether in one’s bank account, a wallet, or a smart contract, are equivalent and fully fungible. There is one monetary base (the Fed’s liabilities) supporting multiple manifestations. Unlike today’s situation where a stablecoin could default or depeg independently, any loss of peg is virtually impossible in the system because of the full-reserve rule and central bank custody.

The Federal Reserve and regulators ensure that no participating stablecoin can be issued beyond its reserves or engage in risky reserve investments. If a stablecoin issuer violates rules, immediate redemption by arbitrageurs will force it back into compliance or out of the system. In effect, the stablecoins become an extension of the monetary base, not a parallel currency. This prevents a scenario where "two dollars" (say, a Federal Reserve issued CBDC dollar and a private stablecoin dollar) diverge in value or credibility, which is a risk that would undermine the dollar's singular role.

Additionally, because the Federal Reserve oversees the reserve backing, it can manage the overall money supply implications. For instance, if demand for digital dollars rises (when people converting bank deposits or cash into stablecoins), those funds end up as reserves at the Federal Reserve, which is part of M0. The Federal Reserve can offset by open market operations if needed to maintain its policy stance.

In other words, monetary policy transmission remains intact. Interest in reserve balances could become a tool to influence stablecoin usage (e.g., if the Federal Reserve pays interest on reserves, perhaps stablecoin issuers could pass some of it to token holders, aligning with policy rates). All of this ensures that even as technology evolves the form of money, the Federal Reserve retains ultimate control over monetary stability and liquidity.

## 5. Theoretical Case

To illustrate the architecture as mentioned earlier, here we consider a user scenario in the following hybrid system.

### *5.1. The scenario*

Alice, an individual, has $500 in a bank account but wants to participate in a DeFi investment. She uses a fintech app to convert $100 into a regulated stablecoin token. Behind the scenes, the issuer takes $100 from Alice via Automated Clearing House (ACH) system from her bank and places $100 into its Fed's reserve account, minting 100 digital tokens in return, which it delivers to Alice's crypto wallet. Alice can then send some tokens to a friend or merchant with ease via a mobile blockchain wallet in seconds (domestic peer-to-peer payment) and deposits the rest into a DeFi lending platform where it earns risk-free interest programmatically. The tokenized money flow is depicted in Fig. 3.

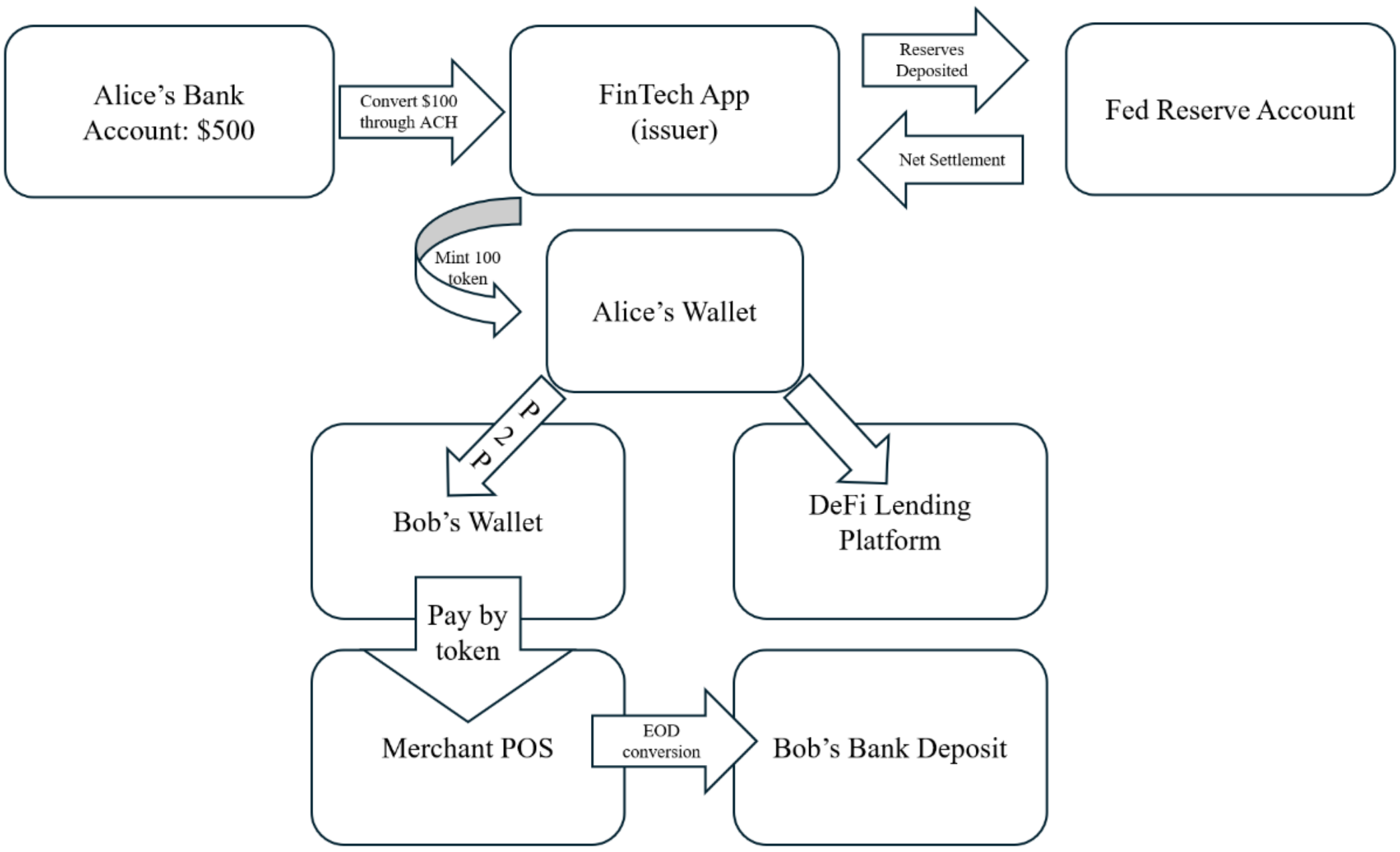

*Alice converts $100 via ACH from her bank into 100 stablecoin tokens, which are fully backed by reserves held at the Federal Reserve. She then uses her crypto wallet for peer-to-peer payments and DeFi investments, while merchants accept the tokens at POS (point-of-sale) and convert them back to USD via an EOD (end-of-day) settlement, which all reconciled with the Federal Reserve in the background.*

**Fig. 3.** Tokenized money flow in the hybrid monetary system.

On the other hand, Bob, a merchant, accepts stablecoin payments from customers (Alice). His point-of-sale (POS) system can convert those tokens to his preferred stablecoin or directly to a bank deposit at the end of day through an integrated service, so there is no price risk in accepting one versus another. If Bob's business would rather keep some revenue in stablecoins, it could invest in a DeFi yield fund first, then redeem to U.S. dollar when needed for supplier payments.

Meanwhile, in the background, the stablecoin issuers and banks cooperate to reconcile net flows through the Federal Reserve's system. If issuer A has net redemptions, it reduces its Federal Reserve's reserves accordingly, and if another issuer B sees net inflows, it then increases B's reserves to balance. The Federal Reserve's balance sheet expands or contracts in tandem, but all dollars remain within the regulated banking perimeter (either as bank deposits or reserve deposits backing tokens). In this way, everyday economic activity can use tokenized money without leaving the safety net of the central bank.

## *5.2. Comparison to the existing system*

This hybrid model absorbs the pros and leaves the cons of the two extremes to create a purely public solution (the Federal Reserve issues retail CBDC to everyone and bans private stablecoins) and a laissez-faire approach (stablecoins can circulate without any special regulation or Federal Reserve's involvement). The purely public model, while guaranteeing uniformity, would suppress private innovation and raise concerns about the government operating retail accounts. The laissez-faire approach, on the other hand, could lead to destabilizing runs or a "wild" situation of unregulated private monies being used under unproper situations.

Note that China's e-CNY pilot program points out both the advantages and downsides of a full-retail CBDC. Its rapid user adoption demonstrates strong demand for a sovereign digital currency, yet early trials highlight trade-offs around privacy (geolocation control, etc.), interoperability with traditional bank accounts, and the high demand for powerful cyber-resilience infrastructures. These practical launch insights from e-CNY suggest that our hybrid design by emphasizing how wholesale-only and optional retail layers can be balanced to mitigate privacy and over-centralization concerns.

Our hybrid approach seeks a balance between keeping what works in the market (speed, innovation, user-driven adoption of stablecoins) and adding the Federal Reserve's imprimatur and oversight to ensure safety and unity. By formalizing stablecoins as part of the system, we also ensure inclusivity. That is, the digital dollar revolution is not limited to crypto enthusiasts but can be accessed via familiar interfaces (your bank or payment app might offer stablecoin services as well as the wallets).

The design is also extensible internationally. If other countries adopt similar frameworks (or if the United States coordinates via standard-setting bodies), cross-border payments could be simplified by bridging CBDCs and stablecoins from different jurisdictions. For instance, a U.S. stablecoin token could be swapped for a digital euro token through automated forex smart contracts, enabling remittances in seconds at an extreme low cost.

## 6. A hybrid system "what-if": SVB bank run and USDC depeg

In March 2023, the crypto market was shaken when USDC broke its $1 peg due to a shock spreading from the traditional banking system. We examine what happened and how the hybrid framework could have improved performance and stability during this crisis. Fig. 4 shows how the price levels changed during the SVB event between March 10 and 15 in 2023.

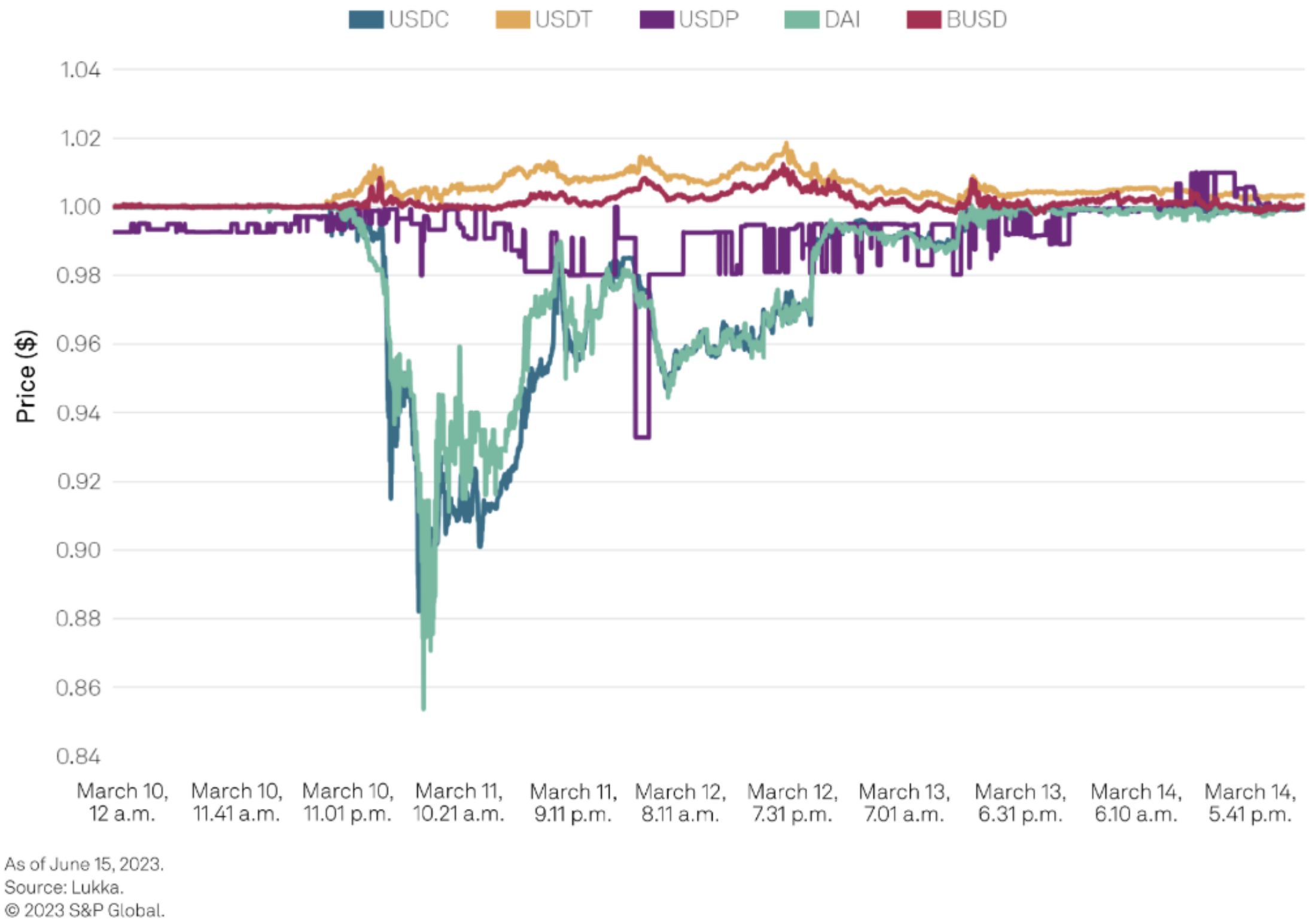

*Price of major USD stablecoins during the Silicon Valley Bank collapse (March 10–15, 2023). USDC (teal) and DAI (green) fell sharply below $1.00 after SVB's failure, while USDT (orange) briefly rose above $1.00. All three recovered to parity within days* [26].

**Fig. 4.** Price levels during USDC SVB event, March 10-15, 2023.

## *6.1. Event background*

On March 10, 2023, Silicon Valley Bank (SVB), a large U.S. bank, collapsed and was taken over by regulators. By the end of 2022, only 6% of the deposit in SVB was insured, and the rest remained uninsured. Circle, the issuer of USDC, disclosed that USD 3.3 billion of USDC's reserve funds were deposited at SVB. This represented about 8% of USDC's total reserve was potentially frozen or lost. Panicked USDC holders rushed to sell or redeem USDC. As a result, USDC's market price plunged to $0.87 to its low on March 11. This depeg was unprecedented for USDC, which historically fluctuated by only pennies.

Other stablecoins also reacted: DAI, with over 50% of its reserved being collateralized by USDC at the time, also dropped to around $0.90. Tether (USDT), which had no exposure to SVB, saw increased demand and traded slightly above $1 (around $1.01 to $1.02). Some trading and DeFi protocols faced chaos due to the unexpected price discrepancies of "supposed-to-be $1" tokens.

## *6.2. Resolution*

The crisis was resolved on March 12 when U.S. authorities (the Treasury, Federal Reserve, and FDIC) announced that all SVB depositors would be made whole on Monday, eliminating the risk that Circle's reserves would suffer a loss from the crisis. Circle also publicly assured that USDC remained redeemable 1:1 and that any shortfall would be covered by other resources if needed. After the news, confidence restored, and USDC's price rapidly climbed back towards peg. By March 13, the price of USDC was around $0.998, and by March 15 it was $1.00 again. In total, USDC spent roughly 3 days under $0.99.

Notably, crypto exchanges and DeFi platforms had to navigate this interim period by ceasing some USDC transactions or updating their codes (for instance, MakerDAO considered emergency measures to prevent DAI from destabilizing further). Ultimately, the situation resolved without any losses to USDC holders, but it exposed a flaw: Stablecoins are only as stable as their backing and the confidence in it. In this case, the backing was entangled with the banking system and reliant on an ad hoc government intervention to avoid a disastrous outcome.

## *6.3. The Diamond-Dybvig model and run equilibrium*

Diamond-Dybvig's framework shows how banks that issue liquid deposits and invest in illiquid assets are subject to self-fulfilling runs arising from the existence of multiple equilibria. In normal times (the no-run equilibrium), depositors believe the bank is solvent and will honor withdrawals, so only those with genuine needs withdraw and the bank can meet demands. But if depositors expect others to run, it creates a panic (the run equilibrium): Everyone tries to withdraw immediately out of fear of being last in line and losing their money.

In SVB's case, we can map the events to a shift from a no-run to a run equilibrium triggered by a shock to expectations:

- Initial state (No-run equilibrium): Before March 2023, SVB depositors generally believed the bank was sound. Even though SVB had large unrealized losses, depositors did not exhibit withdrawal pressure because they expected SVB could meet withdrawals. This aligns with the Diamond–Dybvig no-run scenario: if each depositor expects others not to withdraw en masse, it is rational for them to also leave their funds deposited and avoid causing a run.
- Expectation shock: SVB's March 8 disclosure was the shock that reversed depositor expectations. The announcement of a loss and an emergency capital raise sent a signal that the bank's solvency might be impaired. This quickly became common knowledge among depositors. Large depositors and market commentators expressed concerns about SVB's solvency. In Diamond–Dybvig terms, depositors' belief shifted to "others might run" and therefore accelerated their own withdrawals." Once depositors doubted that SVB had enough liquid assets for everyone, the run equilibrium became the rational expectation. Each depositor had a strong incentive to withdraw immediately because if they waited and others ran, SVB would fail and their uninsured funds could be lost or locked up.

- Run equilibrium dynamics: As the model predicts, once a critical mass began withdrawing, it was optimal for all uninsured depositors to join the run. The fact that about 90% of SVB's deposits were uninsured greatly exacerbated this dynamic. In the Diamond–Dybvig framework, deposit insurance can eliminate the run equilibrium by assuring depositors they will get their money even if the bank fails. But at SVB, most depositors had no such guarantee – their money was at risk if SVB went under. Thus, the rational choice was "withdraw now to mitigate potential loss." Indeed, on March 9 we saw a classic coordination failure: thousands of depositors tried to withdraw approximately $42 billion in one day, far exceeding SVB's available liquidity. Banks operate on a sequential service basis, so those who ran first would withdraw at par value, while late-comers risk being stuck with whatever assets remained. This first-mover advantage is at the heart of the Diamond–Dybvig run equilibrium.

To illustrate with a simplified numerical example: Suppose a bank's long-term assets will be worth $1.00 per $1 of deposits if held to maturity (assuming rational behavior), but only $0.70 per $1 if liquidated early in a fire sale during a run. In a no-run scenario, each depositor can ultimately get $1 back (with interest) by waiting. But in a run scenario, if you wait, you might only get $0.70 (or nothing if others drain the liquidity of the bank first). Facing that prospect, even a rational depositor with no immediate need prefers to withdraw now and obtain the full nominal amount. That logic is self-fulfilling: if everyone follows the same logic, the run occurs and the bank indeed fails, confirming the expectation.

In SVB's run, once the equilibrium shifted, existing market mechanisms were insufficient to arrest the run. Only an external intervention (the government's deposit guarantee on March 12) restored confidence by effectively removing the risk of loss for depositors. This echoes Diamond–Dybvig's prescription that deposit insurance or a lender-of-last-resort can prevent runs by guaranteeing liquidity and breaking the vicious cycle. Notably, after the authorities guaranteed all SVB deposits, withdrawal pressure subsided once loss-given-default was eliminated, since their expectations shifted back to safety.

In summary, the SVB's incident can be seen as a real-world instantiation of the Diamond–Dybvig theory of bank runs. The bank's heavy reliance on uninsured deposits meant the no-run equilibrium was fragile. This underscores that in modern banking, psychology and coordination problems can be just as lethal as bad assets. Even a fundamentally solvent bank can be brought down by a self-fulfilling run if confidence evaporates.

## 6.4. *How could the hybrid system help?*

The hybrid architecture does not promise that stress events will never occur, but it does change both the probability and the propagation of a crisis. To illustrate, consider how the SVB–USDC episode in 2023 would have unfolded had the hybrid system already been in force:

*Prevention of reserve risk:* Under the model, every privately issued dollar token must be collateralized with central-bank liabilities or assets that can be converted to those liabilities at face value within one business day. Had Circle's USD3.3 billion reserve been placed at, or swept daily into, an omnibus account at the Federal Reserve, SVB's failure would still have occurred, but the credit exposure of USDC holders to SVB would have been zero. In other words, the framework eliminates the specific maturity-transformation channel that triggered the de-peg. It does not abolish all sources of volatility (e.g., market sentiment), but it removes the direct link between a commercial-bank insolvency and a stablecoin reserve shortfall.

*Lender of last resort and liquidity:* Full-reserve custody does not preclude temporary liquidity mismatches. For example, if a portion of collateral is held in T-bills that must be settled on a T+1 base, the model can extend standing collateralized credit lines that are parallel to the discount window to licensed issuers. If redemptions surge faster than securities can be settled, the issuer can pledge those same securities to the Federal Reserve and receive immediate settlement balances. This converts what could have become a redemption queue into an operational-timing issue resolved intraday.

*Transparency and trust:* Another improvement is that under a regulated hybrid system, real-time transparency of reserves might be available. If users and traders could see (via publicly verifiable means) that USDC had full backing in safe assets, they might not have panicked as much. In 2023, there was uncertainty and news-driven fear since most USDC holders are unclear about the details of the reserve. The hybrid model could employ blockchain proofs or Fed-published data to show reserve status continuously to avoid any panic caused by uncertainty.

*Federal Reserve/Treasury as the stabilizer:* Under a closer public-private partnership, authorities might directly stabilize a critical stablecoin if needed. Consider if the SVB news had not been resolved so quickly, perhaps regulators might have let uninsured depositors take a haircut, USDC could have remained depegged. In a hybrid system scenario, because of the potential systemic importance, the Federal Reserve and Treasury in coordination with the issuer could act. For example, the Federal Reserve could temporarily swap any reserves that SVB held with its own funds to make the stablecoin safe (effectively a mini rescue for the stablecoin holders, using the issuer's other assets as collateral).

*Interoperability and redundancy:* In our model, hypothetically, if USDC had a problem, people could convert 1:1 into an alternate token through the network, rather than sell at a discount. This mutual fungibility acts as a pressure valve and no single failure would strand users without parity conversion options. It is similar to the deposit insurance in banking, which, even if a bank fails, depositors know their money is safe and effectively interchangeable with money at any other bank up to insurance limits. The hybrid currency network would extend that assurance across digital dollars.

## *6.5. Cross-border AML/KYC*

The hybrid network spans jurisdictions that apply the Financial Action Task Force (FATF) Standards at different levels of maturity. The updated FATF Recommendation in June 2025 now requires originator and beneficiary data (name, address, date of birth) to accompany all peer-to-peer cross-border transfers above 1,000 EUR (or USD), whether processed by banks, virtual asset service providers (VASPs), or token-transfer protocols [27].

Parallel FATF guidance urges supervisors to accelerate licensing of offshore VASPs and to close the stablecoin "gaps", highlighted in the April 2025 Virtual-Assets Targeted Update. All licensed issuers and access gateways participate in an InterVASP-compliant secure channel. For on-chain transfers between self-hosted wallets, gateways attach a hashed Travel-Rule envelope that can be decrypted on regulator demand, meeting the full-information chain test without revealing personal data on-chain.

End-users bind a World Wide Web Consortium (W3C) verifiable credential, issued by a FATF-regulated entity, to their wallet address. When they switch service providers cross-border, the credential is re-used, lowering onboarding friction while preserving the one-time know your customers (KYC) audit trail.

The European Union's General Data Protection Regulation (GDPR), Singapore's Personal Data Protection Act (PDPA), and Switzerland's Federal Act on Data Protection (FADP) all restrict export of personally identifiable information. By encrypting, or zero-knowledge-proofing, the minimum Travel-Rule fields, the network can route compliance packets through regional "compliance enclaves" that keep raw identity data inside the originating jurisdiction. Only salted hash travels with the payment instruction, satisfying both FATF transparency and local privacy statutes.

FATF's June 2025 report notes that stablecoins now dominate on-chain illicit flows, with North Korean-linked addresses laundering an estimated USD51 billion in 2024 [28]. For each gateway, it is important to integrate a real-time blockchain-analytics oracle that: (i) screens inbound/outbound addresses against Office of Foreign Assets Control (OFAC), EU, and UN lists; (ii) assigns a risk score derived from cluster analysis of darknet service exposure; and (iii) auto-quarantines pending enhanced due diligence to prevent the risk of transactions exceeds a pre-set threshold.

## 6.6. *Operational considerations*

Running a single tokenized-dollar ledger across several legal domains is as much an engineering exercise as it is a regulatory one. Even when AML and data-privacy rules are satisfied, the day-to-day requirements of settlement, messaging, and governance still present a distinct layer of operational friction.

The first obstacle is message-format fragmentation. Every token transfer is visible through an on-chain block explorer, giving auditors a single JSON record of hash, amount, and timestamp. That visibility, however, does not eliminate the need to move reserves through legacy real-time-gross-settlement (RTGS) systems that speak different dialects: SWIFT MT-103 in the United States, ISO 20022 pacs.009 in the euro area, and a patchwork of domestic formats elsewhere. Each format expects a field settlement date, purpose code, and fee base that are not natively stored in a token transfer. The network therefore introduces a bridge layer that translates redemption instructions before they hit the chain, mapping them to a canonical ISO 20022 business-purpose code. The result is a single audit trail for supervisors, while issuers avoid maintaining bespoke "explorer-scrape to payment-message" adapters.

A second challenge arises from data-localization mandates. Countries such as India and the United Arab Emirates require that any ledger holding resident data be hosted on servers physically located within their borders. Replicating the chain inside those jurisdictions undoubtedly improves fault tolerance since more copies mean a lower probability of data loss or censorship, but it can also lead to "regulatory forks" if a local supervisor insists on running modified software or censoring sanctioned addresses at the protocol layer. To balance resilience with consensus integrity, the design deploys regulatory tools: read-only validator nodes that mirror the canonical chain in real time yet lack the keys to write blocks. Supervisors gain full visibility and satisfy sovereignty rules, while the authoritative ordering of transactions remains global.

Around-the-clock token liquidity poses a third operational problem, because central-bank RTGS rails still shut down overnight and on weekends. Without careful design, a token redeemed in one time zone could over-draw the issuer's reserve account in another. The hybrid model therefore maintains a prefunded liquidity grid: omnibus reserve accounts in New York (USD), Frankfurt (EUR), and Singapore (SGD) are topped up each business morning, and smart-contract logic caps redemptions to the value already on deposit until the next RTGS window opens. Holders enjoy 24 × 7 convertibility; issuers never incur daylight-overdraft risk.

Finally, the system must accommodate coordinated smart-contract upgrades, whether to patch a security vulnerability or to implement a new regulatory rule. Upgrades follow a two-tier governance process: a two-thirds super-majority of licensed issuers must vote in favor, after which a 24-hour time-lock exposes the finalized code to all home supervisors before activation. Execution occurs through a proxy pattern, so the token balances themselves remain in an immutable contract while the upgradeable logic resides behind a replaceable pointer. That structure ensures continuity of legal title, minimizes downtime, and gives every regulator a chance to veto non-compliant code before it goes live.

Taken together, these mechanisms convert what could be a patchwork of bilateral fixes into a cohesive operational framework. They preserve the security benefits of global replication and on-chain transparency yet keep the system compatible with the heterogeneous settlement rails, sovereignty rules, and supervisory calendars that define today's monetary reality.

## 7. Simulation Results

Instead of relying solely on average regression outputs, we generate thousands of hypotheticals "what-if" price paths that weave together realistic shocks—in trading volume spikes, sudden freezes of reserve funds, and wave-like redemption runs. Each trial randomly combines these stressors in magnitudes and timings drawn from historical behavior, then tracks how the coin's prices depeg from $1 and how quickly to re-peg. By comparing the distribution of outcomes under the existing legend framework versus our hybrid design, where full Fed-backing and emergency liquidity lines mute each shock, we see not just mean effects but the full range of tail risks, worst-case dips, and recovery speeds.

We simulate daily stablecoin price paths over the period November 2019 to April 2025, repeating this process across a large number of trials (N = 20,000). Each trial introduces random shock events (like surging trades, reserve losses, or redemption runs) to stress the peg. The goal is to see how much the hybrid system can dampen price instability compared to the status quo.

### *7.1. Stress scenario design*

We select three shock channels: volume, reserve-loss, and mass redemption, based explicitly on historical stablecoin crises such as the SVB–USDC depeg event in March 2023, where an unexpected reserve impairment led to a significant and rapid market panic. By explicitly modeling these shocks, we replicate the mechanisms through which market participants' confidence can rapidly deteriorate, causing a deviation from the $1 peg.

- Volume shocks: Sudden spikes in trading volume can push the stablecoin's market price away from $1 due to liquidity imbalances or panic trading. Using historical volume-volatility relationships, the simulation model injects days of abnormally high volume and estimates the resulting price impact. Intuitively, on a day when large number of stablecoin users trade urgently, any order imbalances or liquidity shortages could move its price off the $1 peg.

- Reserve shocks (bank failures): This shock captures events like the SVB crisis affecting USDC's reserves. We assign a small probability each day that a portion of a coin's reserves become temporarily inaccessible or impaired (e.g., funds stuck in a failed bank). The magnitude of the reserve freeze is drawn from calibrated distributions. For USDC, we use the actual SVB scenario (about 6% of reserves frozen) as a benchmark. For USDT and DAI, which didn't suffer from the SVB crisis directly, we simulate a comparable what-if reserve freeze (about 1% of reserves) to stress them in a similar way. Such a shock would normally cause a loss of confidence in the stablecoin's backing and drive its market price down (as happened when USDC fell about 10% below peg in March 2023).
- Redemption shocks (bank runs): We also allow for "run on the bank" events when a large share of users suddenly redeem their coins for cash at the same time. In the simulation model, with a small probability (e.g., 0.1% each day), a mass redemption of around 5% of the total stablecoin supply is triggered. This tests the system's ability to handle rapid outflows. Under the current setup, a redemption wave might force the issuer to sell assets quickly or even halt redemptions, potentially causing the price to drop if markets anticipate trouble. The size and frequency of these redemption events are chosen to reflect severe but conceivable bank run scenarios (far larger than normal daily withdrawals, but akin to a panic event).

Each of these shocks is calibrated to real-world data or extreme-case analyses so that the stress scenarios are realistic. For instance, the volume–price impact is based on an empirical regression analysis (as shown in Table 5) linking trading surges to price deviations, and the bank failure probabilities draw on historical bank failure rates, amplified on extreme-volume days to mimic the fact that crises often coincide with high trading activity. By combining volume, reserve, and redemption shocks, the Monte Carlo simulation generates a wide range of crisis narratives, from pure market volatility events to fundamental reserve crises, allowing us to test the stablecoin system under many permutations of stress scenarios.

### *7.2. Current vs. hybrid modeling*

We then impose these same shocks on two alternative system designs to compare their outcomes. In the current system, we assume shocks impact stablecoin price dynamics at full force, as observed historically. In the hybrid system, the effects of shocks are scaled down to reflect the mitigating mechanisms of the hybrid architecture. In practice, this means that for each shock type we apply a fractional multiplier (less than 1) to its impact in the hybrid scenario.

These fractional multipliers represent improvements such as: (i) greater market liquidity and arbitrage efficiency (reducing the price impact of volume surges); (ii) stronger reserve guarantees (largely eliminating the risk that a reserve shock leads to a loss of backing confidence); and (iii) access to central bank liquidity or insurance (meaning even if many users redeem at once, the issuer can meet redemptions without collapsing the peg).

For example, if a sudden sell-off would have caused a 5% dip under current conditions, in the hybrid model it causes only a 1–2% dip because the presence of Federal Reserve backing and real-time oversight reassures arbitrageurs and traders, limiting the price dislocation. Similarly, a reserve shock that might have shaved several percent off the price (due to fears of reserve loss) would be mostly neutralized by the fact that reserves in the hybrid model are safely held at the Federal Reserve (so a bank failure does not actually impair the backing of the coin). And if a mass redemption occurs, the hybrid system allows the stablecoin issuer to draw on its Federal Reserve holdings or an emergency credit line to honor redemptions smoothly, preventing a fire-sale or halt that would otherwise drive the price down.

We do not assume the hybrid system can completely avoid all shocks, but it should substantially weaken them. The exact scaling factors for shock impacts in the hybrid case are chosen based on the design features (e.g., near 0 impact for a reserve freeze, since 100% reserves at the Federal Reserve mean no actual loss occurs, and moderate reductions in volume or redemption impact due to smoother convertibility). We later vary these assumptions to check legitimacy. All formulas for how different shocks translate to price movement in each regime are provided in the Appendix.

After simulating 20,000 trials for each coin (USDT, USDC, and DAI) under both systems, we compute two key performance metrics for every trial: Peak Deviation, which is the worst (maximum) percentage deviation of the stablecoin's price from $1 during the trial, and Off-Peg Days, the total number of days in the trial that the price spent beyond a small tolerance from $1 (i.e., days the coin was significantly "off its peg"). These metrics capture the severity and duration of peg instability, respectively.

By comparing these outcomes between the current and hybrid simulations, we can quantify how much stability the hybrid model adds. For each coin, we take the difference between the current and hybrid system outcomes to see the improvement. For example, if in a given trial, USDC's price fell 8% at worst in the current setup but only 3% in the hybrid, that trial would show a 5 percentage-point reduction in peak deviation for the hybrid system. We aggregate such results across all trials to evaluate the average effect and the variability of that effect across scenarios.

### *7.3. Simulation findings*

The results resoundingly show that the hybrid system confers substantial stability advantages. Table 6 summarizes the average percentage reduction in each metric (with standard deviations in parentheses) achieved by the hybrid design relative to the current design, for USDT, USDC, and DAI.

**Table 6**
Summary of simulation results.

| Coin | Mean improvement in deviation peak ± SD (%) | Mean improvement in off-peg days ± SD (%) | Median improvement in deviation peak (%) | Median improvement in off-peg days (%) |
|---|---|---|---|---|
| USDT | 68.91 ± 28.10 | 57.96 ± 31.65 | 79.99 | 60.00 |
| USDC | 68.87 ± 28.15 | 50.81 ± 31.32 | 80.00 | 50.00 |
| DAI | 68.86 ± 28.16 | 50.17 ± 31.58 | 80.00 | 50.00 |

*This table shows the average improvement that the hybrid model made compared to the legend model across each of the 20,000 simulation runs of the sweeps (from 0 to 1 at a scale of 0.1) for reserve, volume and redemption shocks.*

The key aspects of the results are peak deviation and off-peg days. Peak deviation captures the worst-case scenario of investor loss and market panic, which is an intuitive measure that regulators care deeply about since large, rapid deviations severely disrupt market confidence. Off-peg days measures prolonged instability and recovery speed, reflecting the market's resilience and issuer's capacity to restore confidence quickly after shocks.

Focusing first on peak price deviation, the hybrid system markedly reduces the worst-case peg breaks for all three stablecoins. For USDC, the hybrid system cuts the magnitude of the extreme deviation by roughly on the order of 68.9% on average. For instance, scenarios that would have seen USDC plunge nearly 10% off peg in the current world might only see about a 3.1% dip with the hybrid safeguards in place. USDT and DAI exhibit similarly significant improvements in their worst-day outcomes. In other words, the maximum distance these coins stray from $1 during a crisis is much smaller under the hybrid system.

The fact that all three coins show benefits of peak price deviation and off-peg days, despite their differences in design and backing, underscores that the hybrid system's pillars (full reserves, central bank support, interoperability, etc.) provide a general boost to stability across the board. Equally important, the standard deviations for these improvements are relatively low, indicating that the hybrid advantage is consistent across most trials (not just driven by a few lucky cases).

In every simulation trial, the hybrid configuration either matches or outperforms the current system in limiting peak deviations. We should highlight the fact that there is virtually no scenario where adding hybrid features makes the peak deviation worse. At worst, if no major shock occurs in a simulation trial, both systems show trivial deviation (zero improvement simply because there is nothing to improve upon), and in all non-trivial stress cases, the hybrid system helps contain the price swing. This consistency gives a statistically significant result, given the thousands of simulation trials, the probability that the hybrid's superior performance is due to chance is effectively zero.

*Faster re-pegging:* In addition to reducing how far prices fall from the peg, the hybrid system also reduces how long prices stay away from the peg. Table 7 also reports the average reduction in the number of off-peg days for each coin. The improvements here are striking and practically important that the hybrid model consistently shortens the duration of instability.

For example, suppose that in a severe current-system crisis, USDC might have taken about 10 days to fully recover back to $1. If a hybrid system were in place, that recovery might only occur in about 5 days, which is a 50% reduction in off-peg time (hypothetical figures for illustration). The analysis shows meaningful reductions for USDT and DAI as well.

Essentially, when a shock pushes a stablecoin off parity, the hybrid system enables the coin to return to $1 much more quickly than under current system. This outcome aligns with expectations for a system with reliable backstops. Arbitrageurs in a hybrid system can confidently step in sooner to restore the peg because they know the coin is fully backed and the issuer has liquidity support. Likewise, end-users are less likely to panic-sell in the first place due to greater trust and transparency.

The hybrid network's interoperability (i.e., the ability to swap one stablecoin for another one at 1:1 if an issuer-specific problem arises) also acts as a pressure valve that prevents any single coin's problems from spiraling. The net effect is a faster return to stability. Our simulation results rarely find a hybrid-system coin linger off-peg for long, whereas in the current scenario prolonged deviations are more common whenever a serious shock is present. This improvement in recovery time is as crucial as the reduction in severity, because a short-lived mild depeg is far less damaging than a deep, week-long one.

*Distribution of outcomes:* Fig. 5 provides a visual summary of the stability gains by plotting the distribution (via boxplots) of the peak deviation reduction achieved for each coin. Each boxplot encapsulates the range of outcomes across all simulation trials for that coin's peak peg deviation improvement. The result vividly demonstrates that the entire distribution of improvement is skewed decisively above zero for every stablecoin.

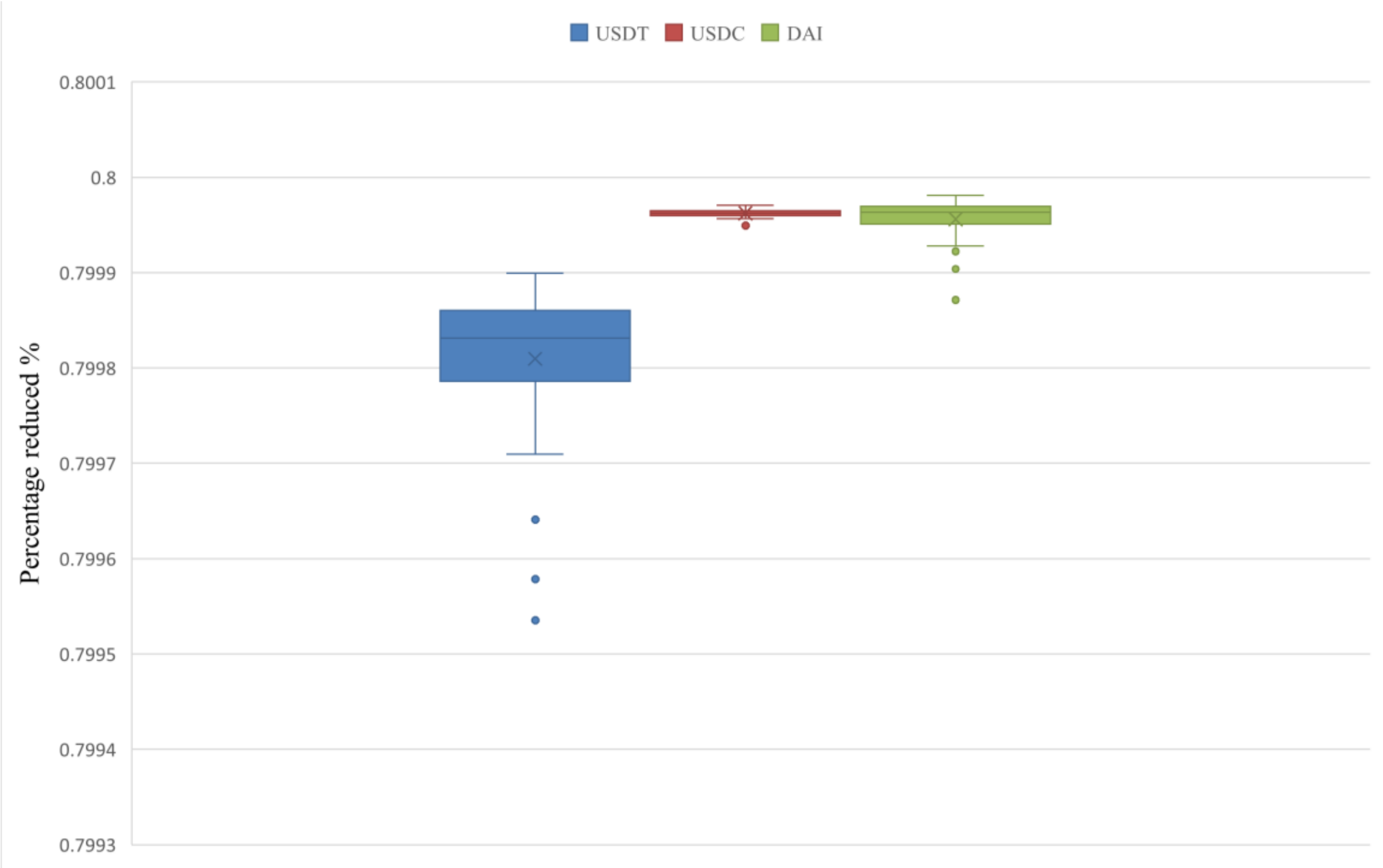

**Fig. 5.** Distribution of peak price deviation reduction under hybrid design.

For USDC, the median reduction in peak deviation is substantially positive, indicating the typical scenario sees a large benefit, and even the lower quartile of outcomes shows a meaningful improvement. Only in the most benign quartile of simulation trials (those with virtually no stress) is the improvement negligible, simply because there was little deviation to begin with.

In particular, the 25th–75th percentile boxes for all three coins lie well above zero, meaning that in at least 75% of random scenarios, the hybrid system delivered a significant reduction in maximum peg loss. The whiskers (covering even the more extreme outliers) remain on the positive side for most cases. It is exceedingly rare to find a trial where the hybrid design underperforms the current system. As noted, the only cases with near-zero difference are scenarios so calm that neither system will experience a notable depeg.

There is no case that shows the hybrid system would ever make a negative impact. The tight clustering of the boxplot distributions (relatively narrow interquartile ranges) also signals that the magnitude of the stability improvement is consistent across different simulations, where the hybrid benefit does not fluctuate wildly or depend on picking a very specific scenario. This consistency provides valid evidence that stability gains are a dependable feature of the hybrid system across a broad spectrum of potential crisis events.

*Parameter sweeps:* Finally, we stress-test our findings against variations in the model's assumptions to ensure the conclusions are held generally. We perform a series of sensitivity analyses, sweeping through the key hybrid effectiveness parameters one at a time. For example, we vary the reserve shock mitigation factor from 0.0 (hybrid almost fully eliminates reserve shocks) up to 1.0 (hybrid only slightly reduces reserve shock impact) and similarly vary the degree to which the hybrid dampens volume and redemption shocks. We also experiment with different overall stress levels, such as making the simulated crises more frequent or severe (e.g., increasing the "extreme day" volume multiplier to simulate a harsher crash scenario).

Across all these variations, qualitative results remain the same: The hybrid system consistently outperforms the current system in maintaining the peg. Naturally, if we assume a less effective hybrid (closer to the current case), the improvement magnitude is smaller. If we assume an optimally effective hybrid, the improvement is larger, but in no case does the hybrid system ever fall below the current system's performance.

Even under very conservative assumptions where, say, the hybrid system only halves or reduces impact by 20%, we still observe significantly lower peak deviations and fewer off-peg days compared to the status quo. This monotonic and reliable improvement across the entire range of parameter sweeps reinforces that our main findings are not an artifact of any single calibration choice.

## *7.4. Simulation results and the classic "narrow banking" argument*

The narrow-banking proposition, foundational in the Diamond–Dybvig framework and highlighted in [29], asserts a clear result: If every deposit is matched one-for-one by a central bank liability, there exists no equilibrium featuring a bank run. A pure narrow banking setup effectively rules out maturity mismatch and systemic panic by construction. Any remaining instability that arises in practice must derive from deviations such as incomplete backing, operational frictions, or imperfect guarantees.

Narrow banking has been considered one of the solutions to deal with financial instability problems in recent years. Using a general equilibrium model with frictions such as limited commitment, moral hazard, and incomplete information, one study shows that deposit insurance, despite its ex-ante distortions, can be more efficient than a hypothetical, perfectly segregated full reserve system [30]. In his framework, narrow banking is only welfare-enhancing when conventional regulations are inefficient or unable to credibly monitor banks; otherwise, it may even reduce social welfare.

The financial risks of tokenized cash, or stablecoin that is fully reserved with high-quality liquid assets, have been examined in [31]. This study provides evidence to demonstrate even fully reserved stablecoins would face short-term liquidity shocks when considering other factors such as empirical run-off rates and liquidity ratios. The study also finds that although these tokens maintain at least twice the amount of high-quality liquid asset (HQLA) reserves, price deviations and temporary dislocations still occur. Our simulation model can extend this static theoretical insight by providing an empirically grounded, quantitative evaluation, producing three key regulatory insights, as shown in Table 7.

**Table 7**
Median peg-stability metrics under baseline probabilities and multiplied reserve-loss stress

(A) Baseline

| | PeakDev legend (%) | PeakDev hybrid (%) | Off days legend | Off days hybrid | ΔPeak Dev (%) | ΔOff days (%) |
|---|---|---|---|---|---|---|
| DAI | 10.37 | 2.08 | 3.0 | 1.00 | 79.97 | 66.67 |
| USDC | 12.22 | 2.44 | 4.0 | 2.00 | 80.01 | 50.00 |
| USDT | 10.18 | 2.04 | 8.0 | 2.00 | 79.96 | 75.00 |
| Average | 10.92 | 2.19 | 5.0 | 1.67 | 79.98 | 63.89 |

(B) 50x multiplier

| | PeakDev legend (%) | PeakDev hybrid (%) | Off days legend | Off days hybrid | ΔPeak Dev (%) | ΔOff days (%) |
|---|---|---|---|---|---|---|
| DAI | 10.80 | 2.16 | 3.0 | 1.00 | 80.01 | 66.67 |
| USDC | 11.33 | 2.27 | 4.0 | 1.00 | 79.97 | 75.00 |
| USDT | 10.48 | 2.09 | 8.0 | 2.00 | 80.02 | 75.00 |
| Average | 10.87 | 2.17 | 5.0 | 1.33 | 80.00 | 72.22 |

*Median peg-stability metrics for DAI, USDC, and USDT are under (A) baseline calibration (mult = 1) and (B) severe reserve-loss stress (mult = 50). PeakDev is the single-largest percentage deviation from par observed along each simulation path. Off-days counts 24-hour periods with a price error ≥ 1 bp. Δ-columns show the percentage reduction achieved by the hybrid full-reserve regime relative to the legacy fractional-custody regime. Each panel summarizes 20,000 simulated paths per coin.*

First, our simulation trials quantify the actual improvement in stability provided by embedding a narrow-banking (full-reserve) arrangement into realistic market scenarios. As shown in Table 7 (baseline scenario), when calibrated to historical probabilities of reserve loss and redemption shocks, median peak deviations from the peg (for DAI, USDC, and USDT) decrease significantly, from approximately 10.92% under fractional reserve custody to around 2.19% under the hybrid, fully reserved model, a 79.98% reduction. Additionally, median annual off-peg days (defined as days with price deviation ≥ 1 basis point) are reduced on average by 63.89%. These results provide clear empirical evidence that even under normal conditions, the hybrid model substantially reduces short-term deviations arising from market frictions or liquidity shocks.

Second, the simulation demonstrates that the hybrid system's stability advantages remain strong even in extreme stress scenarios. As indicated in Table 7 (50x stress scenario), when the historical reserve-loss probability is multiplied by fifty, the hybrid model continues to deliver approximately an 80% improvement in median peak deviations compared to the fractional custody scenario (approximately 10.87% to 2.17%). Furthermore, median off-peg days under severe stress decreased 72.22%. Thus, the hybrid model not only improves stability under typical conditions, but also maintains persistent performance under extraordinary market stress, translating theoretical claims of narrow banking into measurable, practical outcomes.

Third, our simulation results produce concrete benchmarks that regulators can use to establish precise prudential standards. Rather than endorsing an abstract zero-risk narrative, our simulation results quantify the residual peg deviations and off-peg frequencies expected even under full-reserve custody. These results, such as median peak deviations and off-peg days, provide regulators clear, practical metrics to incorporate into frameworks like the U.S. GENIUS Act or the EU MiCA. These benchmarks help define realistic regulatory thresholds for reserve quality, disclosure standards, and liquidity risk management.

Moreover, our findings provide evidence to inform policy discussions on whether and how the public financial safety net that was historically limited to traditional banking could be extended prudently to regulated stablecoin issuers. Specifically, the significant reduction in peak deviations and off-peg durations under the hybrid model highlights that providing stablecoin issuers with controlled and conditional central-bank liquidity access can substantially enhance systemic resilience.

Such an expansion, however, must be accompanied by stringent oversight measures, such as transparent reserve reporting, full-reserve custody requirements, and immediate prudential supervision, to minimize moral hazard risks and maintain market discipline. Our results not only quantify the stability gains achievable through narrow-bank principles, but also contribute to a balanced policy roadmap for integrating stablecoins into the broader monetary safety net framework.

In summary, the simulation model here moves beyond the qualitative narrow-bank narrative to offer clear quantitative metrics for regulators while playing a role in improving stability of the overall safety net. These empirical insights enable regulators to create more precise, enforceable standards that reflect measurable improvements in stability rather than theoretical ideals.

## 8. Challenges and considerations

The implementation of a hybrid monetary ecosystem faces several regulatory and legal challenges. One of the main concerns is the classification and legal status of stablecoins. At this moment, different jurisdictions treat stablecoins variously as e-money, securities, or commodities, leading to regulatory level conflicts.

In the United States, for example, the SEC has indicated that some stablecoins may constitute securities [32], while the CFTC and the Treasury have taken different positions [33]. This lack of definitional consensus complicates efforts to integrate stablecoins into an applicable national structure and slow down innovation due to uncertainty over which agencies have oversight authority.

Second, the integration of private stablecoins with public infrastructure raises questions around consumer protection and systemic risk. A key concern is ensuring that issuers maintain full reserve backing and do not engage in risky lending, as occurred during the 2022 Terra/LUNA collapse [34]. Even fiat-backed stablecoins like USDT have faced criticism over its reserve disclosures.

Newly adopted U.S. legislations, such as the Stablecoin TRUST Act 2022 and the GENIUS Act 2025, seek to establish regulatory clarity by requiring stablecoin issuers to hold reserves in cash or government securities and submit to periodic audits. However, debates remain over whether non-banks should have direct access to Federal Reserve accounts, a critical enabler of the synthetic CBDC model advocated in this paper.

Finally, legal rules must evolve to manage cross-border implications. Stablecoins are borderless, and their integration into payment systems challenges current Anti-Money Laundering/Counter-Terrorist Financing (AML/CFT) standards. The Financial Action Task Force (FATF) has issued guidance that stablecoin arrangements must comply with the Travel Rule and maintain strict identity verification [35].

A hybrid system would require interoperability between national supervisory regimes to avoid arbitrage risks and regulatory loopholes. Failure to harmonize global rules could lead to breakdown of the monetary system, with stablecoins being treated differently across different jurisdictions, ultimately limiting the system's scalability and undermining its trustworthiness.

# 9. Conclusion

The adoption of stablecoins, particularly in emerging markets and decentralized finance, signals a transformative shift in how value is stored, transferred, and accounted for. As proved by both statistical analysis and real-world events like the SVB-USDC de-peg, the existing bifurcation between private digital currencies and fiat money creates latent vulnerabilities. This paper demonstrates that a thoughtfully designed hybrid monetary system, anchored in central bank oversight but powered by private organization innovation, can ease or even resolve these tensions by ensuring peg stability, systemic trust, and technological agility.

Implications for policymakers are substantial. First, enabling regulated stablecoin issuance under a unified legal framework offers a credible route to digitizing the dollar without undermining monetary control. Second, interoperability and composability of these assets could unlock new layers of economic efficiency, from instantaneous multinational payments to programmable decentralized finance. Third, measures must be in place to address concerns such as data privacy and cybersecurity, which are driving risk factors for retail CBDCs and stablecoins. Lastly, the hybrid system needs to align with macroprudential stability objectives of money. It absorbs digital finance into the traditional regulatory perimeter, ensuring that monetary policy, credit creation, and systemic risk remain under the purview of public institutions.

Looking forward, the success of a hybrid monetary ecosystem relies on cross-sector collaboration and regulatory innovation. Governments must build trust-enhancing mechanisms (e.g., real-time reserve transparency), coordinate international standards, and experiment with pilot implementations of synthetic CBDCs. If implemented with care, the hybrid model does not merely accommodate the digital future—it defines and offers a resilient and inclusive framework for 21st-century finance.

## Appendix: Data Analysis Methodologies and Models

Despite its name and design, stablecoins are not necessarily stable in maintaining its prices at times. To better understand the stability of stablecoins in our analysis, we obtained stablecoin data (price, total_volume, market_cap) for DAI, USDC, and USDT from CoinGecko, with daily frequency covering the full sample period. The following macroeconomic variables were sourced from the Federal Reserve Economic Data (FRED) database:

- Federal Funds Effective Rate (DFF) — Daily, 7-Day; Monthly
- M1 Money Stock (M1REAL) — Monthly
- M2 Money Stock (M2REAL) — Monthly
- Median CPI (MEDCPIM158SFRBCLE) — Monthly
- Total Bank Deposits (DPSACBW027SBOG) — Monthly

All datasets were downloaded via direct CSV/Excel exports from FRED. Statistical and econometric analysis was conducted using Python (Pandas, NumPy, Statsmodels) to ensure reproducibility. We performed a variety of statistical analyses and even simulation to better understand the impacts of stability on the current monetary system versus the hybrid one. This section gives just a brief description of the analyses performed. Results and insights will be provided in the following sections.

### *A1. Descriptive statistical analysis*

As a preliminary analysis, we calculate summary statistics for stablecoin prices, trading volumes, and market capitalization. Metrics include mean, median, standard deviation, and range to assess general characteristics and distribution of data. These descriptive measures establish a baseline understanding of central tendency and dispersion of the stablecoins, allowing us to compare the relative scale and variability of each stablecoin's key metrics.

### *A2. Peg stability and volatility analysis*

To evaluate peg stability for each stablecoin, we calculate peg deviation as:

$$Peg\ Deviation = \frac{Price - 1}{1} \times 100\%$$

Volatility is quantified using rolling standard deviation (σ) over a given time (t), according to the formula used in [36]:

$$\sigma_t = \sqrt{\frac{1}{N}\sum_{i=1}^{N-1}(P_{t-i} - \bar{P}_t)^2}\ where\ \bar{P}_t = \frac{1}{N}\sum_{i=1}^{N-1}P_{t-i}$$

and

- $\sigma_t$ is the rolling (realized) volatility at time t

- N is the number of observations in the rolling window (e.g., the past 7 days when computing the 7-day volatility)
- $P_{t-i}$ is the stablecoin closing price on day $t-i$
- $\bar{P}_t$ is the arithmetic mean of the most recent N prices

This peg deviation formula quantifies each day's percentage distance from the $1 target, while the rolling standard deviation captures how price fluctuations evolve over time, highlighting periods of heightened instability.

### *A3. Time series analysis*

The stationarity of stablecoin price series is tested using the Augmented Dickey-Fuller (ADF) test as given in [37]:

$$\Delta y_t = \alpha + \beta t + \gamma y_{t-1} + \sum_{i=1}^{k} \delta_i \Delta y_{t-i} + \varepsilon_t$$

where

- $y_t$ is the price series at time t
- $\Delta$ is the first difference operator
- $\alpha, \beta, \gamma, \delta_i$ are parameters
- $\varepsilon_t$ is white noise

By testing the presence of a unit root (which indicates the time series has a tendency to drift over time; i.e., non-stationary) in time series data, the ADF model determines whether price deviations are mean-reverting (stationary) or follow a random walk, which is critical for understanding peg dynamics. If the time series is found stationary, various standard statistical techniques can be used to model and analyze the data. Otherwise, the data will need to be transformed first before applying those techniques.

### *A4. Regression analysis*

We estimate regression models of peg deviation on volume, market cap, and macroeconomic indicators/variables including interest rates, inflation (CPI), and money supply (M1/M2). This regression framework, similar to [38], quantifies the impact of market activity and broader economic conditions on peg deviations, isolating and identifying which factors most strongly drive stability or divergence within the stablecoins.

### *A5. Monte Carlo Simulation*

In addition to standard statistical analyses as mentioned earlier, we assess the stability benefits of the hybrid monetary system. We applied the simulation procedure and formulas given in [39] to simulate daily price paths $P_t$ for $t = 0, 1, 2, \ldots, T$ corresponding to the period November 2019 to April 2025 (for a total of $T$ days) and perform N independent trials (N = 20,000).

1. Shock channels
    a. Volume shock: The volume shock is being calculated as:
    Volume shocks are drawn from a log-normal distribution calibrated from historical daily trading volumes, capturing realistic extremes (tail events).

    $$V_t \sim LogNormal(\mu_{\ln V}, \sigma_{\ln V})$$

    where $\mu_{\ln V} = \mathbb{E}[\ln V]$ and $\sigma_{\ln V} = Std[\ln V]$ are estimated from historical daily volumes V > 0. The result price impact is

    $$\Delta P_t^{vol} = \beta \frac{V_t}{\bar{V}}$$

    with $\bar{V} = \mathbb{E}[V]$ and $\beta$ from regression displayed in Table 5.
    b. Bank-failure shock
        i. Annual failure probability:

        $$\rho = \frac{\sum_\tau FDIC\ faliures_\tau}{\sum_\tau total\ banks_\tau}$$

        summed over years in [2019, 2025].
        ii. Daily baseline failure probabilities:

        $$p_{base} = 1 - (1 - \rho)^{1/365}$$

        iii. Define the 95th percentile volume threshold $V^{95}$. On "extreme days ($V > V^{95}$), amplified failure risk:

        $$p_t = \min\{1, M_{p_{base}}\}, \quad M \in \{50, 100, 200, 500\}$$

        iv. If a failure occurs ($Bernoulli(p_t) = 1$), draw

        $$L_t \sim LogNormal\ (-3, 0.5)$$

        representing the fraction of reserves frozen. Reserve-shock impact is being calculated as follows:

        $$\Delta P_t^{res} = \gamma L_t$$

    c. Redemption shock:
    Redemption shocks use a 5% mean withdrawal assumption based on industry observations of historical panic events where users collectively redeemed significant portions of stablecoins within short periods (e.g., Tether and USDC redemptions during the Terra/Luna collapse in 2022).

    With probability $p_{red}$ (e.g., 0.001), a mass redemption occurs:

    $$R_t \sim LogNormal(\ln(0.05\overline{MktCap}), 1, )$$

    Centered at 5% of average market capitalization, the redemption impact is calculated as follows:

    $$\Delta P_t^{red} = \delta \frac{R_t}{MktCap_t}$$

    d. Mean reverse and noise:

    $$\Delta P_t^{mr} = \alpha(1 - P_t), \quad \varepsilon_t \sim N(0, \sigma_\varepsilon^2)$$

2. Price dynamics
For each design (Current vs. Hybrid), update daily price by:

$$P_{t+1} = P_t + \alpha(1 + P_t) + \beta \frac{V_t}{\overline{V}} + \gamma L_t + \delta \frac{R_t}{MktCap_t} + \varepsilon_t$$

where

volume = $\beta \frac{V_t}{\overline{V}}$, reserve = $\gamma L_t$, and redemption = $\delta \frac{R_t}{MktCap_t}$.

Let $(\beta, \gamma, \delta)$ denote the Current parameters. For Hybrid, apply fractional scaling:

$$f_\beta,\ f_\gamma,\ f_\delta \quad \in \quad [0, 1]$$

are the Hybrid-fraction parameters to be swept in sensitivity analysis.

3. Parameter calibration
    a. Volume and mean reversion $(\beta, \alpha)$ are taken from OLS regression displayed in Table 5.
    b. Reserve-shock $\gamma$ for USDC
        i. SVB held $3.3 billion of USDC against $56.41 billion total deposits ⟹ L = 3.3/56.41
        ii. Observed peg drop $\Delta P_{obs}$ on 2023-3-10 minus volume impact $\beta \frac{V_t}{\overline{V}}$, gives residual $\Delta_{res}$.
        iii. $\gamma_{current,USDC} = \frac{\Delta_{res}}{L}$
    c. For DAI and USDT, assume 1% frozen at SVB, calibrate $\gamma_{current}$ similarly.
    d. Redemption $\delta$: Current $\delta_{current} = 1.0$; Hybrid $\delta_{hybrid} = f_\delta \delta_{current}$.

4. Performance metrics

    Over each simulated path s:

$$PeakDev_s = \max_{0 \le t \le T} |P_t - 1|$$

    Off-peg days:

$$OffPegDays_s = \sum_{t=0}^{T} 1\{|P_t - 1| > 0.01\}$$

    We report, across trials s = 1…N:

    $\mathbb{E}[Peak\ Dev]$ and the 95th percentile of PeakDev, with mean and median of OffPegDays.

5. Sensitivity analysis

    To demonstrate the legitimacy and valid evidence, we perform three one-dimensional weeps, holding two fractions fixed at baseline $f_\beta, f_\gamma, f_\delta = (0.5, 0.1, 0.2)$:
    a. Reserve Shock

$$f_\gamma \in \{0.0, 0.1, \dots, 1.0\};\ f_\beta = 0.5, f_\delta = 0.2$$

    b. Volume Shock

$$f_\beta \in \{0.0, 0.1, \dots, 1.0\};\ f_\gamma = 0.1, f_\delta = 0.2$$

c. Redemption Shock

$$f_\delta \in \{0.0, 0.1, \dots, 1.0\};\ f_\beta = 0.5, f_\gamma = 0.2$$

For each sweep, we also vary the extreme-volume shock multiplier $M \in \{50, 100, 200, 500\}$ to stress-test under different banking-crash severities.